\documentclass[12pt]{article}
\usepackage{epsfig}
\usepackage{float}
\usepackage{graphicx}
\begin{document}
\rightline{NKU-2016-SF1}
\bigskip

\newcommand{\be}{\begin{equation}}
\newcommand{\ee}{\end{equation}}
\newcommand{\noi}{\noindent}
\newcommand{\ra}{\rightarrow}
\newcommand{\bib}{\bibitem}
\newcommand{\refb}[1]{(\ref{#1})}

\newcommand{\bff}{\begin{figure}}
\newcommand{\eff}{\end{figure}}

\begin{center}
{\Large\bf Phase transitions of   black holes in  massive gravity}

\end{center}
\hspace{0.4cm}
\begin{center}
Sharmanthie Fernando \footnote{fernando@nku.edu}\\
{\small\it Department of Physics, Geology \& Engineering Technology}\\
{\small\it Northern Kentucky University}\\
{\small\it Highland Heights}\\
{\small\it Kentucky 41099}\\
{\small\it U.S.A.}\\

\end{center}

\begin{center}
{\bf Abstract}
\end{center}

In this paper we have studied thermodynamics of a black hole in massive gravity in the canonical ensemble. The massive gravity theory in consideration here  has a massive graviton due to Lorentz symmetry breaking. The black hole studied here has a scalar charge due to the massive graviton and is asymptotically anti-de Sitter. We have computed various thermodynamical quantities such as temperature, specific heat and free energy.  Both the local and global stability of the black hole are studied by observing the behavior of the specific heat and the free energy. We have observed that there is a first order phase transition between small and large black hole for a certain range of the scalar charge. This phase transition is similar to the liquid/gas phase transition at constant temperature for a Van der Waals fluid. The coexistence curves for the small and large black hole branches are also discussed in detail.

\hspace{0.7cm}


%%%%%%%%%%%%%%%%%%%%%%%
\section{ Introduction}

The study of thermodynamics of black holes and phase transitions has gained lot of attention since the seminal work of Hawking and Page on the Schwarzschild-anti-de Sitter black holes \cite{haw} In their seminal work Hawking and Page demonstrated the existence of a phase transition between the Schwarzschild-anti-de Sitter black hole and the thermal AdS space. In the Schwarzschild-anti-de Sitter black hole  there is a minimum temperature a black hole can have. Above this temperature, there could be two black holes; one with large horizon and another with a smaller horizon. The large black hole is thermodynamically stable since it has a positive specific heat capacity while the smaller one is not thermodynamically stable since it  has a negative specific heat capacity. Upon studying the free energy of the black hole spaces and the AdS space, one can see that for smaller temperatures, the thermal AdS space is the thermodynamically preferred  state. On the other hand beyond a certain temperature, the large black holes are preferred. Hence there is first order phase transition between the thermal AdS space and the large Schwarzschild-anti-de Sitter black hole which is knowns as the Hawking-Page phase transition.

Due to the AdS/CFT correspondence \cite{juan}\cite{ed}, the studies of anti-de Sitter black holes have attracted much attention. Studies of phase transitions on anti-de Sitter black holes are done by variety of contexts and by many authors. Thermodynamics of charged black holes in AdS space in various dimensions was done by Chambline et.al. \cite{clif}. Born-Infeld-anti-de Sitter black holes in the grand canonical ensemble were studied by Fernando in \cite{fernando1}. Utilizing the thermodynamic geometry method, Mo and Liu \cite{mo} studied the phase structure of Lovelock AdS black holes in grand canonical ensemble. Geometrothermodynamics technique was used to study thermodynamics of a phantom Reissner-Nordstrom-AdS black hole by Jardim et. al. \cite{jar}. Recently, thermodynamics of AdS black holes have been generalized to the extended phase space where the cosmological constant is treated as the pressure of the black hole. There are many works related to this concept including \cite{cao} \cite{mann} \cite{liu2} \cite{hendi}. One of the first articles to treat the cosmological constant as the pressure in  studying thermodynamics of black holes was by Kaster et. al\cite{kas1} and same authors extended those ideas to AdS black holes in Lovelock gravity in \cite{kas2}.

In general relativity the gravitational field is propagated by the graviton which is massless. It is a  spin two field and has two degrees of freedom. Massive gravity is a deformation of general relativity 
where the graviton is given a mass.  Massive gravity theories have become increasingly popular in the current literature. One  of the reasons to study massive gravity theories is since it cab be  considered a candidate to  explain the acceleration of the universe without introducing  the cosmological constant  or dark energy.  It is speculated that by introducing a mass for the graviton, the gravity can be modified at the infrared in a such a way as to produce the acceleration of the universe.  Existing experimental data put constraints on the mass of the graviton. For example, the recent observation of gravitational waves created by a black hole merger by the Advanced LIGO has put a constraint on the mass of the graviton to be, $m_g < 1.2 \times 10^{-22} eV/c^2$ \cite{ligo}.

The first attempt to include a mass to the graviton was done by Fierz and Pauli in 1939 \cite{pauli}. In such  a theory, the graviton acquires five degrees of freedom. Later, Boulware and Deser showed that Fierz-Pauli theory suffers from ghosts in the non-linear extension \cite{boul}.  Recently, a particular massive gravity theory called dRGT theory  \cite{drgt1}\cite{drgt2} was proposed which  is found to be free from Boulware-Deser ghosts.  There are other models of massive gravity theories free from ghosts in the literature including  DGP model \cite{dgp} and the ``new massive gravity theory''   in three dimensions \cite{town}. There are many  works related to massive gravity in the literature and there is no space to present all here; we will direct the reader to  two excellent reviews on the topic by de Rham \cite{claudia} and  Hinterbichler \cite{kurt} instead.

There are many theories of massive gravity. One of them is a theory with Lorentz symmetry breaking by a space-time dependent condensates of scalar fields. Such scalar fields are called Goldstone fields and are coupled to gravity via non-derivative coupling. When Lorentz symmetry is broken spontaneously, the graviton acquire a mass very similar to the Higgs mechanism.  A  review of Lorentz violating massive gravity theory can be found in \cite{dubo} \cite{ruba2}.

The Lorentz violating theory of massive gravity considered in this paper is given by the following action,
\be \label{action}
S = \int d^4 x \sqrt{-g } \left[ \frac{R}{ 16 \pi}  + \Gamma^4 \mathcal{F}( X, W^{ij}) \right]
\ee
Here the first term is the usual Einstein-Hilbert Lagrangian for general relativity where $R$ is the scalar curvature of the space-time geometry. The second term corresponds to a term of scalar fields $\Phi^{0}, \Phi^{i}$. $\mathcal{F}$  is a function of $X$ and $W^{i j}$ and they are functions of scalar fields $\Phi^{0}, \Phi^{i}$ defined as,
\be
X = \frac{\partial^{\mu} \Phi^0 \partial_{\mu} \Phi^0 }{ \Gamma^4}
\ee
\be
W^{i j} = \frac{\partial^{\mu} \Phi^i \partial_{\mu} \Phi^j}{ \Gamma^4} - \frac{\partial^{\mu} \Phi^i \partial_{\mu} \Phi^0  \partial^{\nu} \Phi^j \partial_{\nu} \Phi^0 }{ \Gamma^8 X}
\ee
$\Gamma$ has dimensions of mass and a perturbative analysis on the theory calculate the value of $\Gamma$ to be in the order of $ \sqrt{ m_g M_{pl}}$. Here  $m_g$  is the graviton mass and $M_{pl}$ the Plank mass\cite{luty}  \cite{dubo} \cite{ruba} \cite{pilo1}. The scalar fields $\Phi^0, \Phi^i$  are responsible for spontaneously breaking Lorentz symmetry where they  acquire a vacuum expectation value.

The paper is organized as follows: in section 2, the black hole in massive gravity is introduced. Thermodynamic stability and phase transitions in black holes are discussed in section 3. Finally the conclusion is given in section 4.

%%%%%%%%%%%%%%%%%%%

\section{ Black holes in massive gravity in AdS space}

In this section we will present the characteristics of the black hole in massive gravity in AdS space. A detailed derivation of such black holes can be found in \cite{tinya} and \cite{pilo}. The metric of such black holes are given by,
\begin{equation} \label{metric}
ds^2 = - f(r) dt^2 + \frac{ dr^2}{ f(r)} + r^2 ( d \theta^2 + sin^2 \theta d \phi^2)
\end{equation}
where,
\begin{equation} \label{fr}
f(r) = 1 - \frac{ 2 M} { r} - \gamma \frac{ Q^2}{r^{\lambda}} - \frac {\Lambda r^2}{3}
\end{equation}
In the paper by Bebronne and Tinyakov\cite{tinya} the black hole derived did not have a cosmological term in the $f(r)$. The integration constants in the derivation was chosen such that $\Lambda=0$. However, one can pick the integration constants such that $ \Lambda \neq 0$ which can lead  to the function given in eq$\refb{fr}$. In this paper, we chose a negative cosmological constant to represent AdS space. 

\begin{figure} [H]
\begin{center}
\includegraphics{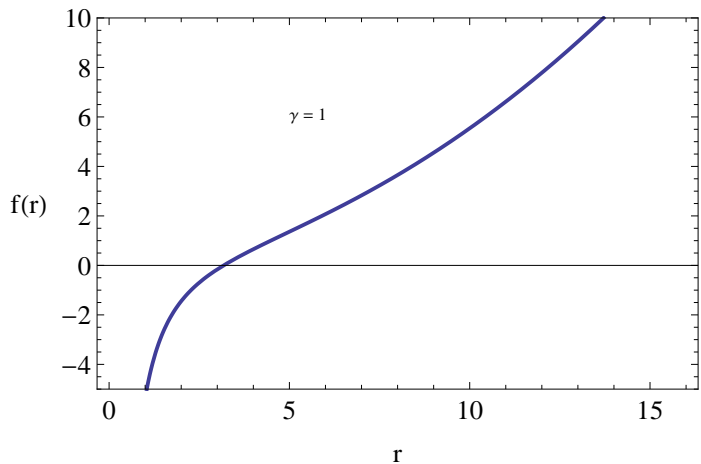}
\caption{The figure shows  $f(r)$ vs $r$ for $ \gamma =1$. Here $M = 1.55$, $ \Lambda = -0.148$, $ Q = 1.82$, and $\lambda =1.6$.}
\label{f1}
 \end{center}
 \end{figure}

The integration constant   $Q$ represents a scalar charge related to massive gravity. The parameter $\gamma$ can take $\pm 1$ which lead to different geometries.
The scalar fields $ \Phi$ described in the introduction are given as,
\be
\Phi^0 = \Gamma^2 \left( t + \beta(r) \right); \hspace{1 cm} \Phi^i =   \Gamma^2 x^i
\ee
where
\be
\beta(r) = \pm \int \frac{ dr} { f(r)} \left[ 1 - f(r) \left( \frac{ \gamma Q^2 \lambda(\lambda-1)}{ 12 m_g^2 b^6}\frac{1}{ r^{\lambda+2}} + 1 \right)^{-1} \right]^{1/2}
\ee
Here $m_g$ is the mass of the graviton and $\lambda$  is an integration constant and is positive. The cosmological constant $\lambda$ is related to the parameter $b$ as, $\Lambda = 2 m_g^2 ( 1 - b^6)$.
\noi
The function $\mathcal{F}$ for this particular black hole solution is given by,
\be
\mathcal{F} = \frac{ 12 b^6} { \lambda} \left( \frac{ 1}{ X} + q_1 \right) - \left( q_1^3 - 3 q_1 q_2 - 6 q_1 + 2 q_3 - 12\right)
\ee
where,
\be
q_n = Tr( W^n)
\ee
When $ \gamma = 1$, the geometry is very similar to the Schwarzschild-AdS  black hole with a single horizon. The function $f(r)$ for this case is given in Fig.$\refb{f1}$.  The integration constant $\lambda$ could have any value greater than zero. However, when $\lambda <1$, the third term in $f(r)$ dominates at large distances and the ADM mass of such solutions will be divergent. For $\lambda >1$,  the metric approaches the standard Schwarzschild-AdS black hole with a finite mass $M$. Hence for the rest of the paper, we will consider $\lambda >1$.

The horizon radius for the black hole in massive gravity is larger than the one for the Schwarzschild-AdS black hole. When $\lambda \ra \infty$,  $ r_h \ra r_{Sch-AdS}$ which is the value for the Schwarzschild-AdS black hole.

\begin{figure} [H]
\begin{center}
\includegraphics{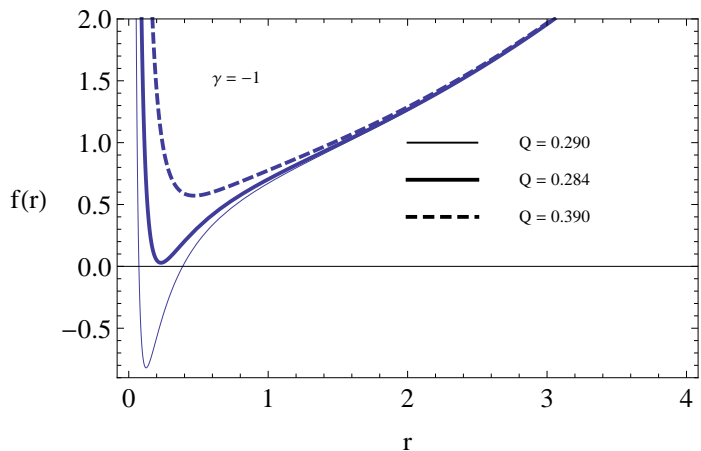}
\caption{The figure shows  $f(r)$ vs $r$ for $ \gamma =-1$. Here $M = 0.25, \Lambda = -0.371$ and $\lambda = 1.835$.}
\label{f2}
 \end{center}
 \end{figure}

When $\gamma = -1$, the geometry is similar to the well known Reissner-Nordstrom charged black hole. There could be two, one(extreme black hole)  or no horizons depending on the parameters  chosen for the theory.  The function $f(r)$ for this case is given in Fig.$\refb{f2}$.

When the mass of the black hole $M$ is,
\be
M_{extreme} =  \frac{ 3 r_e \lambda - r_e^3 ( 2 + \lambda) \Lambda}{ 6 ( \lambda - 1)}
\ee
the horizons merge and form an extreme black hole. The radius of the extreme black hole, $r_e$, and $Q_e$ are related as,
\be
Q_e^2 =\frac{ r_e^{\lambda} ( 1 - \Lambda r^2)}{ ( \lambda -1)}
\ee
For $M > M_{extreme}$,  there will be two horizons. For $M < M_{extreme}$, no horizons exists and  there will be a naked singularity.

Before we complete this section we want to mention some of the work in the literature related to the above black hole. Quasi normal modes of the black holes with zero cosmological constant were studied for  the scalar and the Dirac field by Fernando \cite{fernando2}\cite{fernando3}. Thermodynamics and phase structure of black holes with $\Lambda =0$ were studied by Capela and Nardini \cite{capela}. Phase transitions of the above black hole with $\Lambda =0$ were studied by Mirza and Sherkatghanad \cite{mirza}. External stability of spherically symmetric solutions of Lorentz breaking massive gravity was studied in \cite{and}.

%%%%%%%%%%%%%%%%%%%%%%%%%%%%%%%%%%%%%

\section{Thermodynamic stability of the black hole in the canonical ensemble}

In this section we will explore thermodynamics of the black holes considering the black hole as a closed system, i.e, the scalar charge $Q$ is kept constant. Hence the black hole is considered to be in the canonical ensemble. The behavior of the thermodynamical quantities differ considerably for $\gamma =1$ and $\gamma= -1$. Hence we will discuss the two separate cases  for each quantity in the following sections.

%%%%%%%%%%%%%%%%%%%%%%%%%%%%%%%%%%%%%

\subsection{ Temperature}

\begin{figure} [H]
\begin{center}
\includegraphics{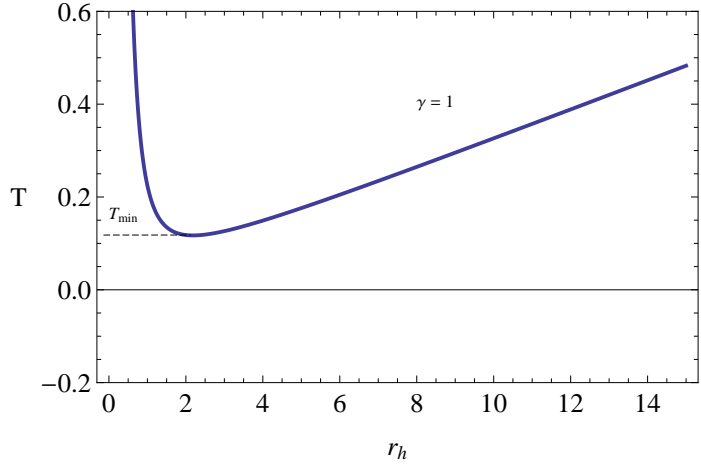}
\caption{The figure shows  $Temperature$ vs $r_h$ for $\gamma= 1$. Here $Q = 1.22, \Lambda =-0.4$ and $\lambda = 1.955$.}
\label{temp1}
 \end{center}
 \end{figure}

\begin{figure} [H]
\begin{center}
\includegraphics{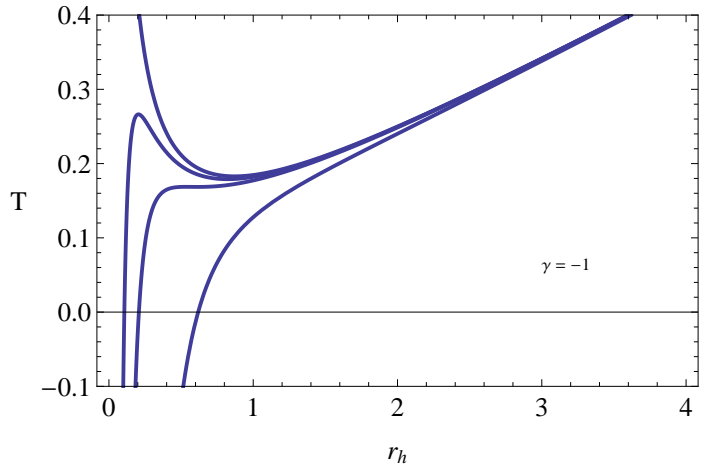}
\caption{The figure shows  $Temperature$ vs $r_h$ for $\gamma =-1$. Here $\Lambda = -1.32$ and $\lambda = 1.55$. The top graph is at $Q=0$ and then the rest are plotted for increasing charge with $Q =0.24,0.41,1.14$.}
\label{temp2}
 \end{center}
 \end{figure}

The Hawking temperature of the black hole is given by,
 \be \label{temp}
 T_H =  \frac{ 1}{ 4 \pi}  \left| \frac{ df(r)}{ dr} \right|_ { r = r_h} = \frac{1}{ 4 \pi} \left(\frac{ 2 M}{ r_h^2} + \frac{ \gamma Q^2 \lambda}{ r_h^{ \lambda + 1}}   - \frac{ 2 \Lambda r_h}{ 3}\right)
 \ee
 Here $ r_h$ is the black hole horizon. Since $ f(r_h) = 0$,
 \be \label{mass}
 M = \frac{r_h}{2} - \frac{ \gamma Q^2}{ 2 r_h^{( \lambda -1)} }-  \frac{ r_h^3 \Lambda}{6}
 \ee
 When $M$ in eq.$\refb{mass}$ is substituted to eq.$\refb{temp}$,  the temperature can be written as a function of $(r_h, Q, \Lambda, \lambda)$ as,
\be
T = \frac{ 1 }{ 4 \pi} \left( \frac{1}{ r_h}  -  r_h \Lambda   + \frac{ \gamma  ( \lambda -1)Q^2}{ r_h^{ \lambda +1} } \right)
\ee
 The first law of the black hole  is given by
 \be \label{flaw}
 dM = T dS + \Phi dQ
 \ee
 Hence the entropy can be computed as,
 \be
 S = \int^{r_h}_0 \left(\frac{ \partial M}{ \partial r_h } \right)\frac { 1}{ T} dr  = \pi r_h^2
 \ee
The thermodynamic potential corresponding to the scalar charge $Q$ is computed as,
\be
\Phi = \left ( \frac{ \partial M} { \partial Q } \right) _S =  \left ( \frac{ \partial M} { \partial Q }\right) _{r_{h}} = - \frac { \gamma Q}{ r_h^{ \lambda -1}}
\ee
The temperature is plotted as a function of  $r_h$ in Fig$\refb{temp1}$ and Fig.$\refb{temp2}$. In Fig$\refb{temp1}$, the temperature is plotted for $\gamma =1$ where only one horizon exists. There is a minimum for the temperature ($T_{min}$) which is positive. The minimum divides the black holes to small and large.  Above the minimum temperature, small and large black holes coexists at all temperatures. This behavior of the temperature is very similar to the behavior of the temperature of the Schwarzschild-AdS black hole \cite{haw}.

In Fig.$\refb{temp2}$, the temperature is plotted when $\gamma =-1$ where two horizons can exists. Various values of $Q$ is employed to plot the behavior of the temperature. The top curve corresponds to $Q=0$ which corresponds to the Schwarzschild-AdS black hole. There, the temperature has a minimum (similar to the $\gamma=1$ case described above).  When $Q$ is increased, the temperature  has two turning points (minima and a maxima). Further increasing the temperature brings these two turning points to one point leading to an inflection point. After that the curve does not have any turning points.

\begin{figure} [H]
\begin{center}
\includegraphics{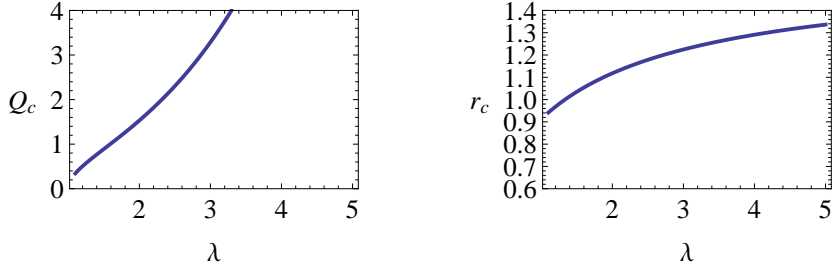}
\caption{The figure shows  $Q_c$  and $r_c$ vs $\lambda$  for $\Lambda = - 0.4$}
\label{qc}
 \end{center}
 \end{figure}

The critical charge $Q_c$ is the value of $Q$ at which the $T$ vs $r_h$ curve has an inflection point. At this point,
\be
\frac{ \partial T}{ \partial r_h } =  \frac{ \partial^2 T}{ \partial r_h^2 } =0
\ee
After some algebra, $Q_c, r_c$ and corresponding $T_c$ can be derived to be,
\be
r_c =  \sqrt{\frac {\lambda}{ -\Lambda ( 2 + \lambda)}}
\ee
\be
Q_c =  \sqrt{ \frac{2 r_c^{\lambda} }{ (2  +\lambda) ( \lambda^2 -1)}}
\ee
\be
T_c = \frac{  \sqrt{ - \lambda ( 2 + \lambda) \Lambda}}{ 2 \pi  ( 1 + \lambda)}
\ee
The plots for $Q_c$ and $r_c$ are given in Fig$\refb{qc}$. One can observe that both $Q_c$ and $r_c$ increase with the parameter $\lambda$.

When $Q < Q_c$, there can be three branches of black hole solutions (small, intermediate and large).

%%%%%%%%%%%%%%%%%%%%%%%%%%%%%%%%%%%%%%%

\subsection{Local stability}

\begin{figure} [H]
\begin{center}
\includegraphics{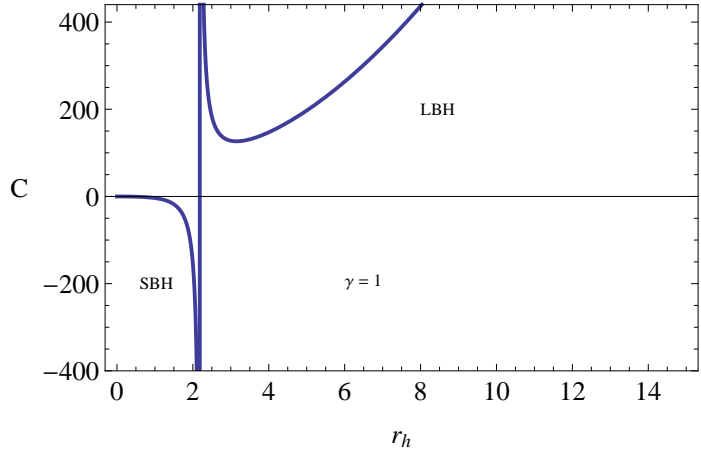}
\caption{The figure shows  $C_Q$ vs $r_h$ for $\gamma =1$. Here $Q = 1.22, \Lambda =- 0.4$ and $\lambda = 1.955$.}
\label{spe1}
 \end{center}
 \end{figure}

The local stability of the black hole can be studied by computing the specific heat of the black hole at constant charge $Q$ as,
\be
C_Q = \left(\frac{ \partial M}{ \partial T_H } \right)_Q = \left(\frac{ \partial M}{ \partial r_h } \right)_Q  / \left(\frac{ \partial T_H}{ \partial r_h } \right)_Q
\ee
After some algebra, $C_Q$ becomes,
\be \label{sh}
C_Q = \frac{ -2 \pi r_h^2 \left( r_h^\lambda + \gamma Q^2 ( \lambda -1) - r_h^{ 2 + \lambda} \Lambda \right)}{ \left(r_h^{\lambda} + \gamma Q^2 ( \lambda^2 -1) + r_h^{\lambda + 2} \Lambda \right)} = \frac{ A(r_h, Q)} { B(r_h, Q)}
\ee
The black hole will be locally stable if the specific heat is positive. First we will study the local stability of the black hole for $\gamma =1$. The graph for $C_Q$ vs $r_h$ is plotted in Fig.$\refb{spe1}$. Here,  $C_Q < 0$ until it reaches a singular point. The singular point of $C_Q$ is where the temperature $T$ has a minimum ($T_{min}$) in Fig$\refb{temp1}$.  $C_Q >0$ after the singular point. Hence, the small black holes(SBH) are unstable and the large black holes (LBH) are stable.
 
\begin{figure} [H]
\begin{center}
\includegraphics{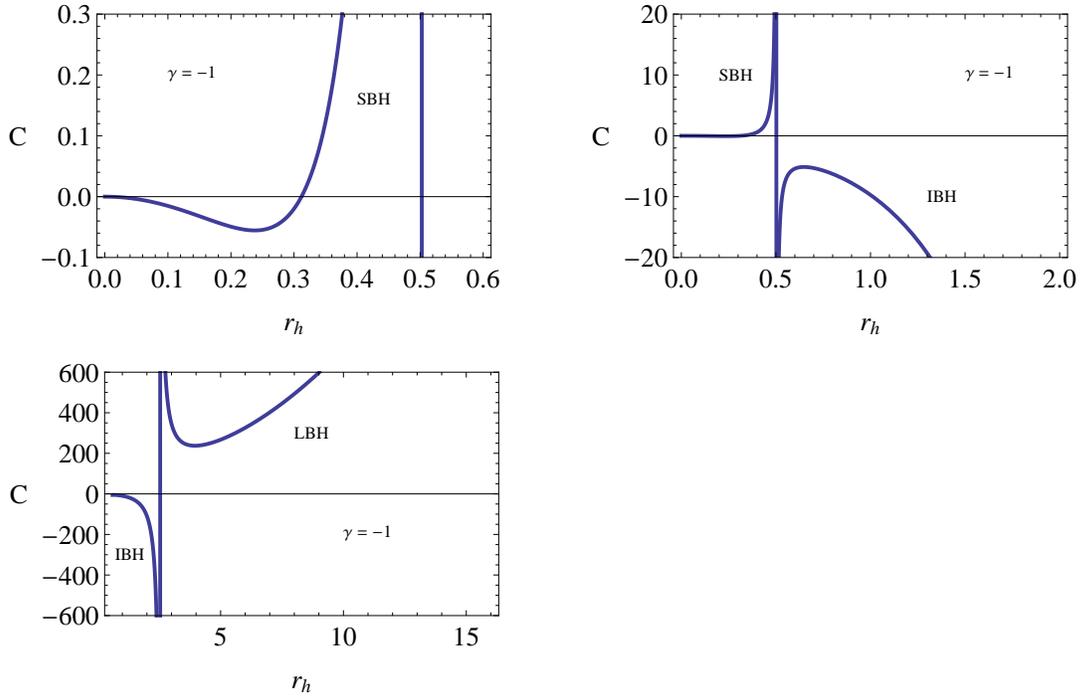}
\caption{The figure shows  $C_Q$ vs $r_h$ for $\gamma = -1$. The three graphs shows various values of $r_h$. Here $Q = 1.2, \Lambda =- 0.154$ and $\lambda = 3.04$}
\label{spe2}
 \end{center}
 \end{figure}

When $\gamma =-1$, there are two singular points for $C_Q$ as shown in Fig.$\refb{spe2}$. These two corresponds to the minima and maxima of the temperature shown in Fig.$\refb{temp2}$.  When $r_h$ is small $C_Q =0$ and $T=0$ coincides. Since $T<0$ is unphysical, we will describe the behavior beyond $T >0$. From  the graph, $C_Q >0$ for small black holes(SBH), hence they are stable. For intermediate black holes (IBH) $C_Q <0$ hence they are unstable. For large black holes (LBH) $C_Q>0$ hence they are stable.

%%%%%%%%%%%%%%%%%%%%%%%%%%

\subsubsection{Critical exponent for $C_Q$}

From the previous section, it is clear that the specific hear capacity $C_Q$ diverges at various values for the horizon radius; we will call these points critical points. Since the critical point is a mark of the divergence of the heat capacity, it is important to understand the nature of this singular behavior. The nature of the thermodynamic systems around the critical points are usually studied by the computing the critical exponents.  Here, to understand the behavior  of $C_Q$ near the critical points ($r =r_i$)  we will calculate the critical exponents for $C_Q$. We will follow the method presented by Banerjee and Roychowdhury \cite{ban} to compute the critical exponents.

The divergence of $C_Q$ at $r=r_i$ can be described by
\be
C_Q \approx | T(r_h) - T(r_i)|^{-\alpha}
\ee
In order to compute $\alpha$, we will choose a point in the infinitesimal neighborhood of $r_i$ as,
\be \label{delta}
r_h = r_i + r_i \bigtriangleup
\ee
Here $ |\bigtriangleup| << 1$. Now one can define a new parameter $\epsilon$ such that,
\be \label{epsilon}
\epsilon = \frac{ T(r_h) - T(r_i)}{ T(r_i)}
\ee
Here $|\epsilon| << 1$. $T(r_h)$ can be expanded in the neighborhood of $r_i$ for constant charge $Q$ as the following:
\be \label{taylor}
T(r_h) = T(r_i) + \left. \left(\frac{ \partial T}{ \partial r_h } \right)_Q \right|_{r = r_i} ( r - r_i) + \frac{1}{2} \left. \left(\frac{ \partial^2 T}{ \partial r_h^2 } \right)_Q \right|_{r = r_i} ( r - r_i)^2
+ ........
\ee
Since $\left(\frac{ \partial T}{ \partial r_h } \right)_Q=0$ at $ r=r_i$, eq.$\refb{taylor}$ becomes,
\be \label{taylor2}
T(r_h) = T(r_i) + \frac{1}{2} \left. \left(\frac{ \partial^2 T}{ \partial r_h^2 } \right)_Q \right|_{r = r_i} ( r - r_i)^2
\ee
where we have omitted the higher order terms. Combining eq.$\refb{delta}$ $\refb{epsilon}$ and eq.$\refb{taylor2}$, one can write
\be \label{delta2}
\bigtriangleup = \sqrt{\frac{ \epsilon}{ \mathcal{K}_i}}
\ee
with
\be
\mathcal{K}_i = \frac{ r_i^2}{ 2 T(r_i)} \left. \left(\frac{ \partial^2 T}{ \partial r_h^2 } \right)_Q \right|_{r = r_i}
\ee
After substituting $r_h = r_i + r_i \bigtriangleup$ to $C_Q$ in eq.$\refb{sh}$ and expanded, one arrive at the singular part of  $C_Q$ as,
\be \label{sh2}
C_Q = \frac{ A(r_h, Q)} { \bigtriangleup B'(r_i,Q)}
\ee
We have obtained the numerator of the above equation by expanding $B(r_i + r_i \bigtriangleup, Q)$ around $ r = r_i$ only to first order of $\bigtriangleup$ as the following,
\be
B( r_i + r_i \bigtriangleup, Q) = B(r_i, Q) + r_i^{\lambda} \bigtriangleup \left(  \lambda + r_i^2 \Lambda(2 + \lambda)\right)
\ee
Since $B(r_i, Q) =0$, $B'(r_i, Q)$ becomes,
\be
B'(r_i,Q) = r_i^{\lambda} \left(  \lambda + r_i^2 \Lambda(2 + \lambda)\right)
\ee
Now, substituting for $\bigtriangleup$ from eq.$\refb{delta2}$ in eq.$\refb{sh2}$, one arrive at,

\be \label{sh3} C_Q =  \left\{ \begin{array}{ll}

\frac{\mathcal{A}_i}{\sqrt{- \epsilon}}  & \mbox{if $\epsilon < 0$}\\

\frac{\mathcal{A}_i}{\sqrt{\epsilon}}  & \mbox{if $\epsilon > 0$}\\

\end{array}
\right. \ee

where
\be
\mathcal{A}_i = \frac{ \sqrt{\mathcal{K}_i} A( r_h, Q)}{ B'(r_i)}
\ee
If we combine the eq.$\refb{sh3}$ into a single expression to describe the singular nature of $C_Q$ at $ r = r_i$, we arrive at,
\be
C_Q =  \frac{ \mathcal{A}_i} { \sqrt{\epsilon}} = \frac{\mathcal{A}_i \sqrt{T(r_i)}}{ | T(r_h) - T(r_i)| ^{1/2}}
\ee
From the above equation, the critical exponent can be derived to be $ \alpha = \frac{1}{2}$.

%%%%%%%%%%%%%%%%%%%%%%%%%%%%%%%%%%%%%%%%%%%%%%%

\subsection{Global stability}
Global stability of black holes can be understood by studying the free energy of the black hole. In the canonical ensemble, the black hole  remain in thermal equilibrium at constant temperature with the heat reservoir (or a bath of radiation) while its energy is allowed to fluctuate. Here the scalar charge $Q$ is kept constant. This type of ensemble is described by the Helmholtz free energy given by,
\be
 F = M - T S = \frac{r}{4} - \frac{ \gamma Q^2} {4  r^{\lambda -1}} ( 1 + \lambda) + \frac{ r^3 \Lambda}{ 12}
\ee
The derivative of the free energy is given by,
\be
\frac{ d F}{ dr_h} = \frac{ r^{-\lambda}}{4} \left( r^{\lambda} + \gamma Q^2 ( \lambda^2 -1) + r^{2 + \lambda} \Lambda \right)
\ee
It is clear that the specific heat diverges when $\frac{dF}{dr_h} =0$. 

Now we will discuss the global stability of the black hole for $\gamma = \pm 1$ values as follows:

%%%%%%%%%%%%%%%%%%%%%%%%%%%%%%%%%%%%%%%%%%%%

\subsubsection{ Phase transition for $\gamma =1$}

\begin{figure} [H]
\begin{center}
\includegraphics{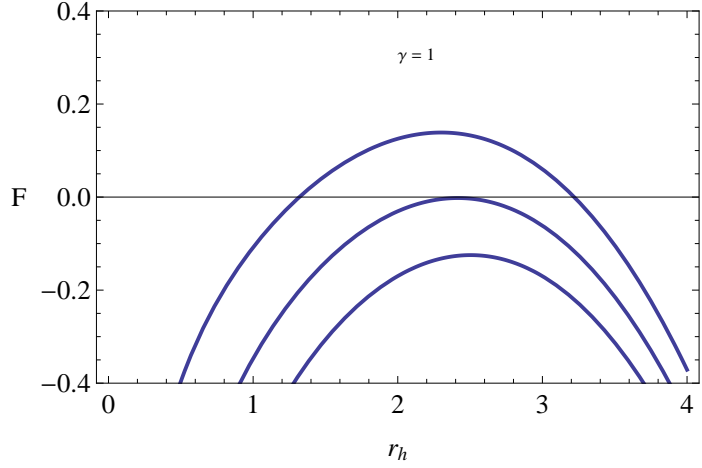}
\caption{The figure shows  $F$ vs $r_h$ for $ \gamma = 1$. Here $\Lambda =-0.23$, $\lambda = 1.615$ and $Q = 0.72, 0.94, 1.1$ from top to the bottom}
\label{free1}
 \end{center}
 \end{figure}

\begin{figure} [H]
\begin{center}
\includegraphics{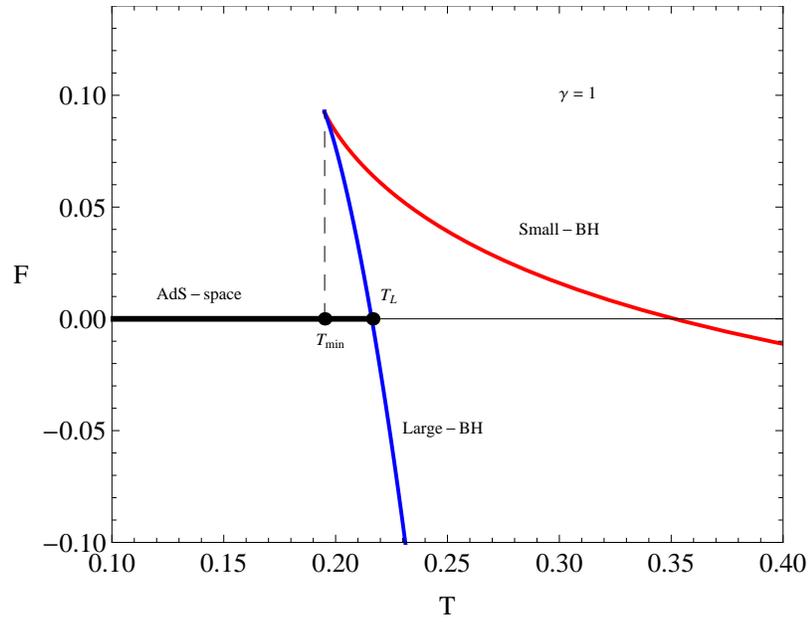}
\caption{The figure shows  $F$ vs $T$ for $\gamma = 1$. Here $Q = 0.268, \Lambda =-1.45$ and $\lambda = 1.388$}
\label{free2}
 \end{center}
 \end{figure}

\begin{figure} [H]
\begin{center}
\includegraphics{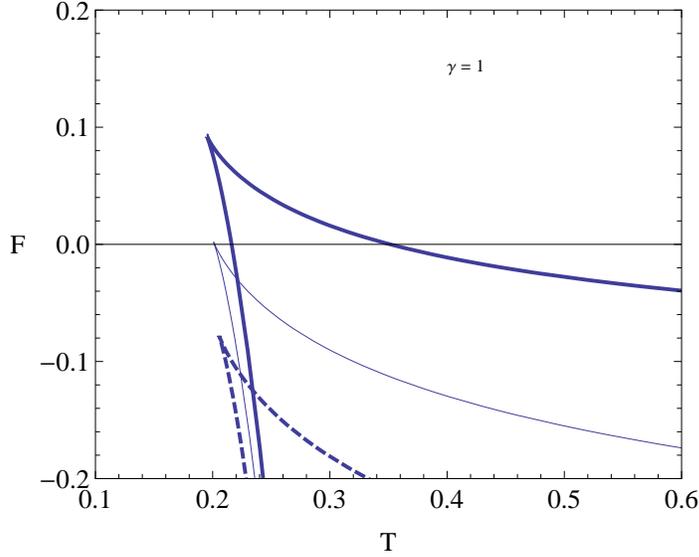}
\caption{The figure shows  $F$ vs $T$ for $ \gamma = 1$. Here $\Lambda =-1.45$, $\lambda=1.388$ and $Q = 0.268, 0.466, 0.59$ from top curve to the bottom respectively}
\label{free3}
 \end{center}
 \end{figure}

First we will plot $F$ vs $r_h$ as given in Fig.$\refb{free1}$. For small $Q$ the free energy is positive for some values of $r_h$. When $Q$ is increased, there is a critical value of $Q_c$ that makes $F$ negative completely. $Q_c$ and the corresponding $r_c$ are given by,
\be
Q_c = \sqrt{ -\frac{2 \Lambda r_c^{3 + \lambda}}{ 3 \lambda ( 1 + \lambda)}}
\ee

\be
r_c = \frac{ 3 \lambda}{ \sqrt{ -\Lambda ( 2 + \lambda)}}
\ee
When $Q > Q_c$,  $F$ is negative. To further analyze the global stability of the system, $F$ is plotted against $T$ in Fig.$\refb{free2}$ and Fig.$\refb{free3}$.
First we will look at Fig.$\refb{free2}$ where $Q < Q_c$.  $T_{min}$  in the figure is the same as the minimum temperature in Fig.$\refb{temp1}$. From $0< T< T_{min}$, no black hole can exists. Hence for that range of the temperature, the thermal AdS state is preferred. When $ T  > T_{min}$, there are two branches of black holes. The upper branch (in red) corresponds to the small black holes and they have negative specific heat. Hence these black holes are thermodynamically unstable and cannot be in thermal equilibrium  with a heat reservoir. The lower branch (given in blue) corresponds to large black holes and they have positive specific heat. Hence they are locally thermodynamically stable.

Beyond $T = T_{min}$, the free energy of both the small and the large black holes are positive. Hence the thermal AdS space is the globally preferred thermodynamic state until $T = T_L$. After that, the large black holes have negative free energy and are thermodynamically preferred. At $T = T_L$, the horizon radius $r_L$ can be found by solving $ F =0$.  From these observations, one can see that there is a first order phase transition between thermal AdS space and large black holes in massive gravity. This phase transition is very similar to the one between thermal AdS space and the large Schwarzschild-AdS black holes \cite{haw}.

When $Q$ is increased, the free energy of both kinds of black holes become more negative as shown  in Fig.$\refb{free3}$. However, the large black holes have smaller  free energy in all cases and are preferred over small black holes for all values of $Q$.

%%%%%%%%%%%%%%%%%%%%%%%%%%%%%%%

\subsubsection{ Phase transition for $\gamma =-1$}

\begin{figure} [H]
\begin{center}
\includegraphics{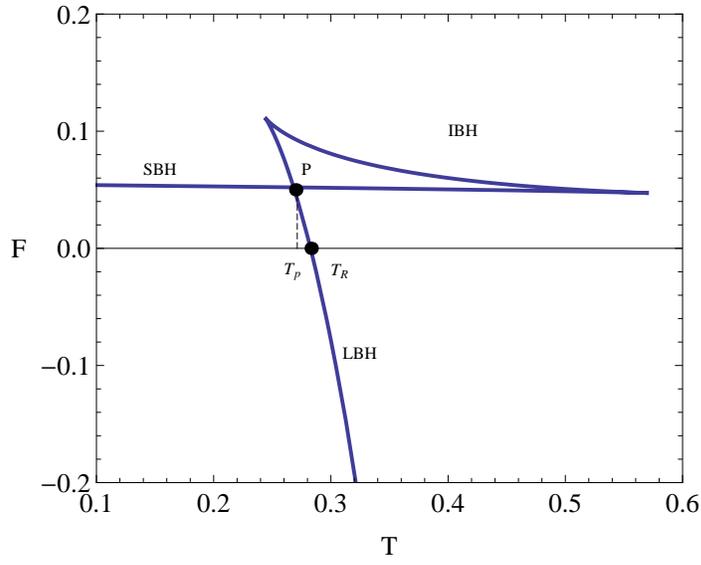}
\caption{The figure shows  $F$ vs $T$ for $\gamma = -1$. Here $ Q = 0.052, \Lambda =-2.36$ and $\lambda = 2.036$.}
\label{free5}
 \end{center}
 \end{figure}

When $\gamma =-1$, there are three branches of black holes; small, intermediate and large. As discussed in section(3.2), the small and large black holes are locally stable and the intermediate black holes are locally unstable. Now one can study the global stability of the black holes by observing the free energy. In Fig.$\refb{free5}$ the free energy is plotted as a function of $T$ for small charge. Notice that for small change an extreme black hole is possible.

Now, from $T =0$ to point $P$ in the graph, the system will go along the small black holes branch. One may wonder why the thermal AdS space with zero energy is not chosen as the preferred state. When $ T=0$, the corresponding black holes are extreme and if we lower the charge below $Q_e$ then there won't be any black holes; instead there will be naked singularities. Hence in this case, we will consider the extreme black hole as the ground state. This is  similar to the argument used when the free energy was discussed for the Reissner-Nordstrom-anti-de Sitter black hole in the canonical ensemble where   extreme black holes exists \cite{clif}. Since from  $T=0$ until point $P$, the lowest free energy state is the small black hole branch, it is taken as the preferred thermodynamical state. At point $P$ and beyond  the large black hole branch is preferred since the free energy is smaller compared to the small black holes. Hence at point $P$, there is a small-black hole/large-black hole  phase transition. The temperature corresponding to the point $P$ is given as $T_p$.

Since the small and large black holes have different radius, there is discontinuity in the black hole area at point $P$.  Since the entropy of black holes are given by the area, there is discontinuity of the entropy at point $P$ leading to a release of latent heat. One can take the difference in the radius of the black holes, $ \sigma = r_h^{LBH} - r_h^{SBH}$ as an order parameter for the phase transition. The phase transition at point $P$ is first order. 

The first order phase transition can be more systematically described by studying the Helmholtz free energy $F$. Using  the definition of $F$ and  combining it with the first Law equation in eq$\refb{flaw}$, $dF$ can be written as,
\be
dF = - S dT + \Phi dQ
\ee
From above, one can write the two equations,
\be \label{entropy}
S = -\left(\frac{ \partial F}{ \partial T } \right)_Q
\ee
and
\be
\Phi = \left(\frac{ \partial F}{ \partial Q } \right)_T
\ee
In a first order phase transition, the first derivative of $F$ is discontinuous at phase boundary. For the black hole free energy in Fig.$\refb{free5}$ one can see that in fact  $\left(\frac{ \partial F}{ \partial T } \right)_Q$ is discontinuous at point $P$.  Hence from eq$\refb{entropy}$, the entropy is discontinuous as well.

\begin{figure} [H]
\begin{center}
\includegraphics{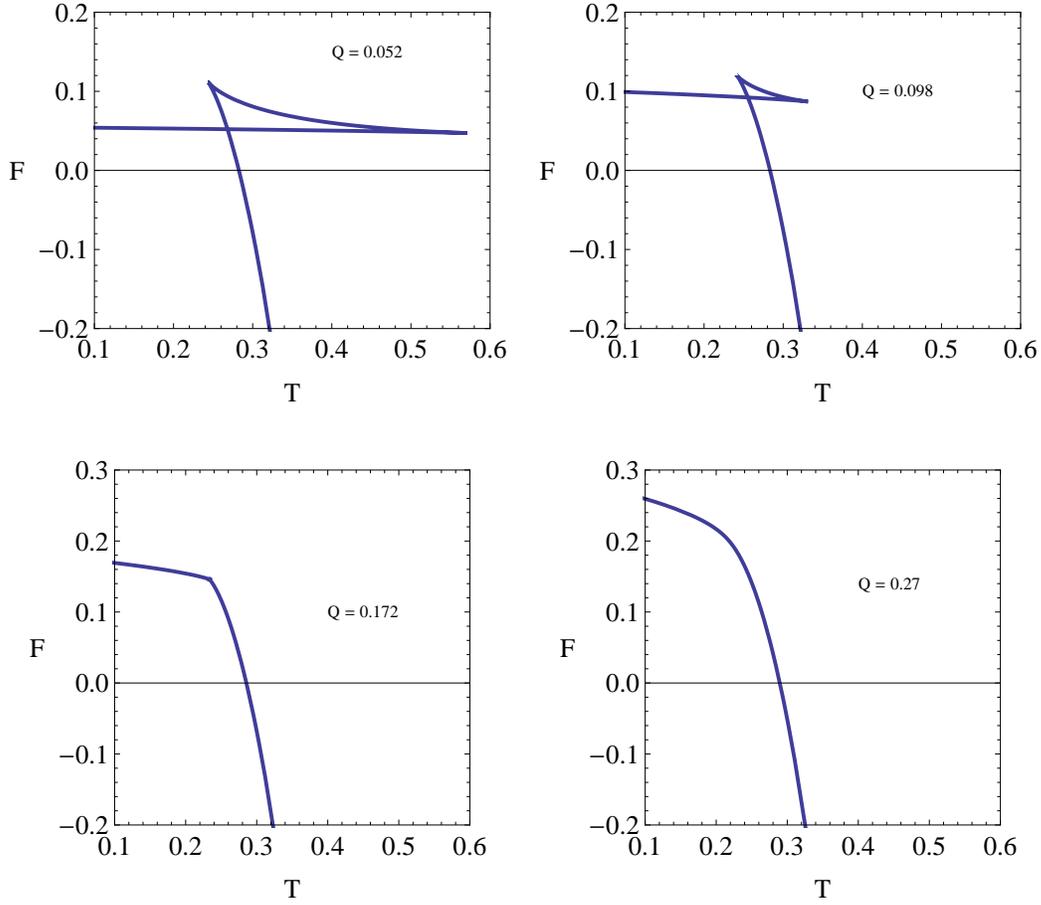}
\caption{The figure shows  $F$ vs $T$ for $\gamma = -1$. Here $\Lambda =-0.4$ and $\lambda = 1.955$.}
\label{free4}
 \end{center}
 \end{figure}

The phase transition between the small and the large black hole  is similar to the liquid/gas phase transition at constant temperature for a Van der Waals fluid when the volume of the gas suddenly changes at the critical pressure \cite{hill}. In order to make comparison with the AdS black hole considered  in  this paper with the liquid/gas phase transitions, one has to identify the physical parameters in this paper with the corresponding ones in the Van der Waals system as given in the table below:

\begin{center}
\begin{tabular}{|l|l|l|l|l|r} \hline \hline
 Fluid/Gas system  & AdS black hole in canonical ensemble \\ \hline

Pressure(P) &  Temperature(T)  \\ \hline
Volume(V) &   Horizon radius($r_h$)    \\ \hline
Temperature(T) & Scalar charge(Q)   \\ \hline
\end{tabular}
\end{center}
When $Q$ is increased until it reachers $Q_c$, the kink in the free energy plot will disappear. This is demonstrated in Fig$\refb{free4}$. When the kink vanishes, the phase transition also vanishes and so does the order parameter $\sigma$.

\begin{figure} [H]
\begin{center}
\includegraphics{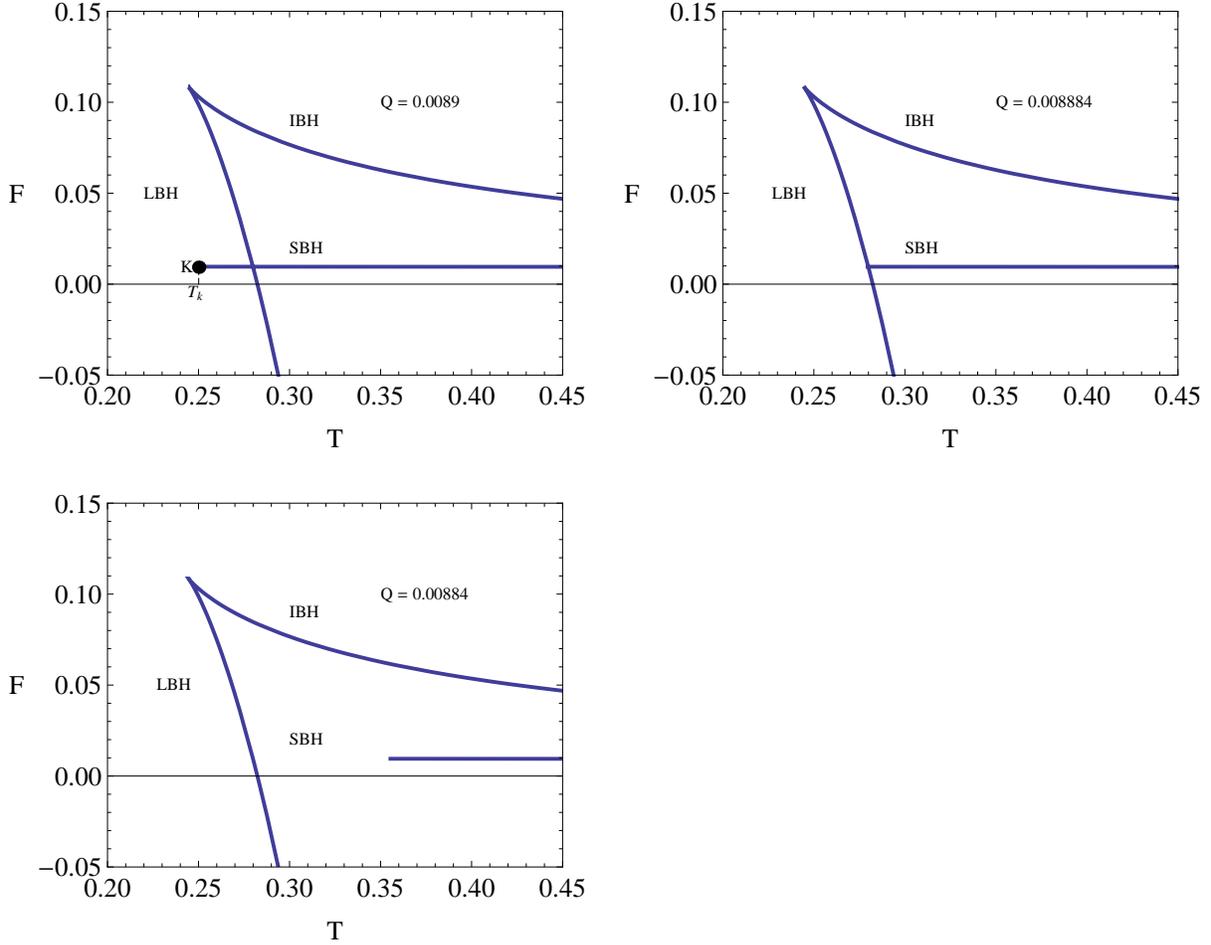}
\caption{The figure shows  $F$ vs $T$ for $\gamma = -1$. Here $\Lambda =-0.4$ and $\lambda = 1.955$.}
\label{freecritical}
 \end{center}
 \end{figure}

The phase transition between small black holes and large black holes does not exist for small charge. When the charge gets smaller, the small black hole branch in the $F$ vs $T$ graph begins its existence at larger temperature and, when $Q$ gets smaller the curve moves to the right as is evident from Fig.$\refb{freecritical}$. Hence when $Q$ is smaller than a specific value, the large black hole and the small black hole branches cease to have common free energy values. Therefore there won't be phase transitions below the specific $Q$ value. The preferred thermodynamical state will be large black hole below this particular charge value. We will name this value of $Q$ as $Q_t$.

%%%%%%%%%%%%%%%%%%%%%%%%%%%%%%%

\subsection{ Coexistence curves for the phase transition}

In this section we would like to investigate the coexistence curve for the phase transition between the small and the large black holes. Coexistence for the Reissner-Nordstrom-AdS black holes have been studied in \cite{liu3} and for the charged black holes in $f(R)$ gravity have been studied in \cite{gu}.

\begin{figure} [H]
\begin{center}
\includegraphics{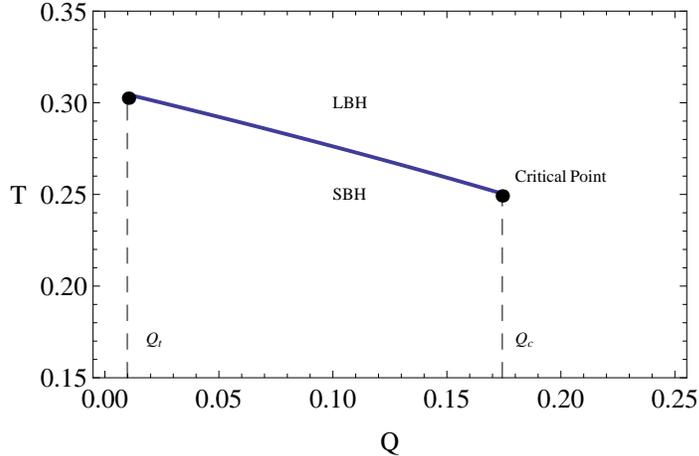}
\caption{The figure shows  coexistence curve for small and large black holes during the phase transition.  Here $T$ vs $Q$ is plotted for  $\Lambda = - 2.8$.  For this case $ Q_t = 0.009808$ and $Q_c = 0.172516$. The phase transitions exists only between $Q_t$ and $Q_c$.}
\label{coexitence}
 \end{center}
 \end{figure}

We will use $r_1$ and $r_2$ for the radius of the small and the large black holes respectively. Let us depict the free energy for small black holes as $F_1$ and for large black holes as $F_2$. Hence both of $F_1$ and $F_2$ can be written as,
\be
F_1 = \frac{r_1}{4} + \frac{ Q^2} {4  r_1^{\lambda -1}} ( 1 + \lambda) + \frac{ r_1^3 \Lambda}{ 12}
\ee
\be
F_2 = \frac{r_1}{4} + \frac{  Q^2} {4  r_2^{\lambda -1}} ( 1 + \lambda) + \frac{ r_2^3 \Lambda}{ 12}
\ee
Let us also depict the temperature of the black holes as $T_1$ and $T_2$ where each is given as,
\be
T_1 = \frac{1}{ 4 \pi} \left(\frac{ 2 M}{ r_1^2} - \frac{  Q^2 \lambda}{ r_1^{ \lambda + 1}}   - \frac{ 2 \Lambda r_1}{ 3}\right)
\ee
\be
T_2 = \frac{1}{ 4 \pi} \left(\frac{ 2 M}{ r_2^2} - \frac{  Q^2 \lambda}{ r_2^{ \lambda + 1}}   - \frac{ 2 \Lambda r_2}{ 3}\right)
\ee
Now, the black holes undergo phase transition along a coexistence curve. Hence along this curve, the free energy of both black holes are the same implying $F_1 = F_2$. This relation can be represented as,
\be \label{comfree}
\frac{r_1}{4} + \frac{ Q^2} {4  r_1^{\lambda -1}} ( 1 + \lambda) + \frac{ r_1^3 \Lambda}{ 12} = \frac{r_1}{4} +  \frac{  Q^2} {4  r_2^{\lambda -1}} ( 1 + \lambda) + \frac{ r_2^3 \Lambda}{ 12}
\ee
Also, during the phase transition, the temperature both black holes are the same leading to $T_1 = T_2 = T$. This relation can be expressed as,
\be \label{comtemp}
\frac{1}{ 4 \pi} \left(\frac{ 2 M}{ r_1^2} - \frac{ Q^2 \lambda}{ r_1^{ \lambda + 1}}   - \frac{ 2 \Lambda r_1}{ 3}\right) =\frac{1}{ 4 \pi} \left(\frac{ 2 M}{ r_2^2} - \frac{  Q^2 \lambda}{ r_2^{ \lambda + 1}}   - \frac{ 2 \Lambda r_2}{ 3}\right)
\ee
One can also add both temperatures and obtain another formula as,
\be \label{addtemp}
 2T = \frac{1}{ 4 \pi} \left(\frac{ 2 M}{ r_1^2} - \frac{  Q^2 \lambda}{ r_1^{ \lambda + 1}}   - \frac{ 2 \Lambda r_1}{ 3} + \frac{ 2 M}{ r_2^2} - \frac{  Q^2 \lambda}{ r_2^{ \lambda + 1}}   - \frac{ 2 \Lambda r_2}{ 3}\right)
 \ee
Since the equations are quite complicated for a general value of $\lambda$, we will study the coexistence curves for $\lambda = 2$ case here. Now, one can introduce two parameters, $ r_1 + r_2 = x$ and $ r_1 r_2 = y$ and rewrite the above equations $\refb{comfree}$, $\refb{comtemp}$ and $\refb{addtemp}$ as,
\be
- 9 Q^2 + 3 y + x^2 y \Lambda - y^2 \Lambda =0
\ee
\be
Q^2 ( x^2 - y^2) - y^2 - y^3 \Lambda =0
\ee
\be
- Q x^3 + 3 Q x y + x y^2 - x y^3 \Lambda -  8 \pi T y^3 =0
\ee
One can solve the above three equations to obtain,
\be
T = \frac{ 1 }{ 8 \pi} \sqrt{\left( \frac{ -64 \Lambda}{ 3} + \frac{ 128 Q (-\Lambda)^{\frac{3}{2}}}{ 3 \sqrt{3}} \right)}
\ee
The relation between $T$ and $Q$ for the coexistence is plotted in Fig.$\refb{coexitence}$.  

%%%%%%%%%%%%%%%%%%%

%%%%%%%%%%%%%%%%%%%

\section{ Conclusion}

In this paper, we have studied thermodynamics of black holes in massive gravity in anti-de Sitter space. The system was studied in the canonical ensemble where the scalar charge due to the graviton is kept constant. Thermodynamic 	quantities such as temperature, specific heat and the free energy were computed to study stability of such black holes.

For $ \gamma = 1$ values in the theory, the massive gravity black hole has similar behavior as the Schwarzschild-anti de Sitter black hole. The temperature has a minimum value.  The specific heat is singular at this temperature and is negative for small black holes and positive for small black holes. Hence small black holes are unstable and large black holes are stable. We observed that the thermal anti-de Sitter space is  globally preferred thermodynamic state for small temperature  and beyond $T = T_L$
 the large black holes are preferred. Hence, there is a first phase transition between the thermal AdS space and the large black holes.

 For $\gamma = -1$, the behavior of the black hole temperature is much more complex. For certain values of $Q$, the temperature could have a minimum and a maximum. When $Q$ is increased the maxima and the minima coincide to give a point of inflection. For $\gamma = -1$, there are three branches of black holes; small, intermediate and large. We observed that for certain values of the charge $Q$, there is a first order phase transition between small and large black holes. This behavior is very similar to the liquid/gas phase transition at constant temperature for a Van der Waals fluid. Only difference is that for the system of black holes, the temperature of the black hole is like the pressure and the horizon radius is analog to the volume. We also studied the coexistence curves for this phase transitions between the small and the large black hole and showed that there is range of the charge $Q$ where both kind of black holes can coexist.
 
As future work related to the work presented here, it would be interesting to study thermodynamics of the  extended phase space where the negative cosmological constant is considered as the pressure given by the relation $ P = - \frac{\Lambda}{ 8 \pi}$. 

\vspace{0.5 cm}

%%%%%%%%%%%%%%%%%%%%%

%%%%%%%%%%%%%%%%%%%%%%

\end{document}